\documentclass[11pt]{article}
\usepackage{epsfig,moriond}
\usepackage{graphicx}
\usepackage{amsmath}
\usepackage{amssymb}
\usepackage{slashed}

\def\be{\begin{equation}}
\def\ee{\end{equation}}
\def\bea{\begin{eqnarray}}
\def\eea{\end{eqnarray}}

\begin{document}
\vspace*{0cm}
\title{CGC phenomenology at RHIC and the LHC}

\author{ J.~L. Albacete}

\address{Institut de Physique Th{\'e}orique, CEA/Saclay, 91191 Gif-sur-Yvette cedex, France\\
URA 2306, unit\'e de recherche associ\'ee au CNRS}

\maketitle
\abstracts{I present a brief review of the recent phenomenological analyses of RHIC data based on the the Color Glass Condensate, including the use of non-linear evolution equations with running coupling. In particular, I focus in the study of the total multiplicities in Au+Au collisions, and in the single inclusive and double inclusive forward spectra in d+Au collisions. Predictions for the LHC are also discussed}

At high energies, QCD scattering enters a novel regime governed by large gluon densities and coherent, non-linear phenomena, including {\it saturation} of the hadronic wave functions, known as the Color Glass Condensate (CGC) (see e.g., the review \cite{Gelis:2010nm} and references therein). 
Nuclear collisions performed at RHIC provide a good opportunity to explore the CGC regime, since the gluon densities in a nucleus are already large even at moderate energies. 
The recent calculation of running coupling corrections \cite{Balitsky:2006wa,Kovchegov:2006vj,Gardi:2006rp,Albacete:2007yr} to the BK-JIMWLK evolution equations of the CGC allows for a good quantitative description of several experimental measurements, thus reducing the degree of modelization required in phenomenological studies. Here I present a brief review of the phenomenological analyses of RHIC data based on the use of Balitsky-Kovchegov equation  \cite{Balitsky:1995ub,Kovchegov:1999yj} including running coupling correction (rcBK) for the description of the small-$x$ degrees of freedom of the wavefunction of the colliding nuclei. We also discuss predictions for the LHC. 

The rcBK equation for the small-$x$ evolution of the dipole scattering amplitude reads  
\begin{equation}
  \frac{\partial {\cal N}(x,r)}{\partial\ln(x_0/x)}=\int d^2r_1\
  K^{{\rm run}}(r,r_1,r_2) \left[{\cal N}(x,r_1)
+{\cal N}(x,r_2)-{\cal N}(x,r)- {\cal N}(x,r_1)\,{\cal N}(x,r_2)\right]\ ,
\label{bk1}
\end{equation}
where $r_i$ refers to the dipoles tranverse sizes and $x_0$ is the starting point for the evolution. The running coupling kernel $K^{{\rm run}}$ is evaluated according to Balitsly's prescription \cite{Balitsky:2006wa}. In the analyses presented below, the rcBK equation is supplemented with MV initial conditions at the starting evolution point, $x_0$, $\mathcal{N}_{F(A)}(r,x_0)=1-\exp\left[ -\frac{r^2\,Q_{s0}^2}{4}\,\ln\left(\frac{1}{\Lambda\,r}+e\right)\right]$, with $Q_{s0}$ the initial quark (gluon) saturation scale and $\Lambda\!=\!0.241$ GeV.

\section{Multiplicity densities in nucleus-nucleus collisions a RHIC and the LHC.}
The total number of particles produced per unit rapidity in RHIC Au+Au collisions turned out to be significantly lower than predicted assuming incoherent superposition of nucleon-nucleon scattering, signaling the importance of coherence effects. Such reduced multiplicities are interpreted in the CGC as a consequence of a reduced flux of scattering centers, i.e. gluons, entering the collision due to saturation effects in the wave function of the colliding nuclei. Thus, the final number of produced particles rises proportional to the number of gluons in the incoming wavefunction, whose growth can be described by the rcBK equation. Even if its applicability to nucleus-nucleus collisions is not completely justified, such idea is realized in the $k_{T}$-factorization framework, as proposed in \cite{Kharzeev:2001yq}, where the multiplicity distributions can be written as \cite{}:
\begin{equation}
\frac{dN_{ch}}{dy\, d^2b}=C\frac{4\pi
  N_c}{N_c^2-1}\int\frac{d^2p_t}{p_t^2}\int^{p_t}\,d^2k_t\,
\alpha_s(Q)\,
\varphi\left(x_1,\frac{\vert\underline{k_t}+\underline{p_t}\vert}{2}\right)
\varphi\left(x_2,\frac{\vert\underline{k_t}-\underline{p_t}\vert}{2}\right), 
\label{ktfact}
\end{equation}
where $p_t$ and $y$ are the transverse momentum and rapidity of the
produced particle, $x_{1,2}\!=\!(p_t/\sqrt{s})\,e^{\pm y}$. $\varphi(x,\underline{k}_t)=\int \frac{d^2r}{2\pi\,r^2}e^{i\,k_t\cdot r}\,\mathcal{N}(x,r)$ is the nuclear unintegrated gluon distribution. The lack of
impact parameter integration in this calculation and the gluon to
charged hadron ratio are accounted for by the constant $C$, which sets the
normalization. As shown in Fig. 1, the use of Eq. (\ref{ktfact}) together with rcBK equation for the small-$x$ dynamics of the nuclear ugd provides  \cite{Albacete:2007sm} a good description of the energy and pseudorapidity dependence of data  \cite{Back:2004je} for the multiplicity densities in Au+Au collisions at the highest collision energies at RHIC. With all the parameters in the calculation constrained by RHIC data, the extrapolation to Pb+Pb collisions at the LHC is now completely driven by the small-$x$ dynamics, yielding $\frac{dN^{Pb-Pb}_{ch}}{d\eta}(\sqrt{s}\!=\!5.5\,\mbox{TeV},\eta=0)\sim1290\div1480$.

\begin{figure}
\begin{center}
\epsfig{figure=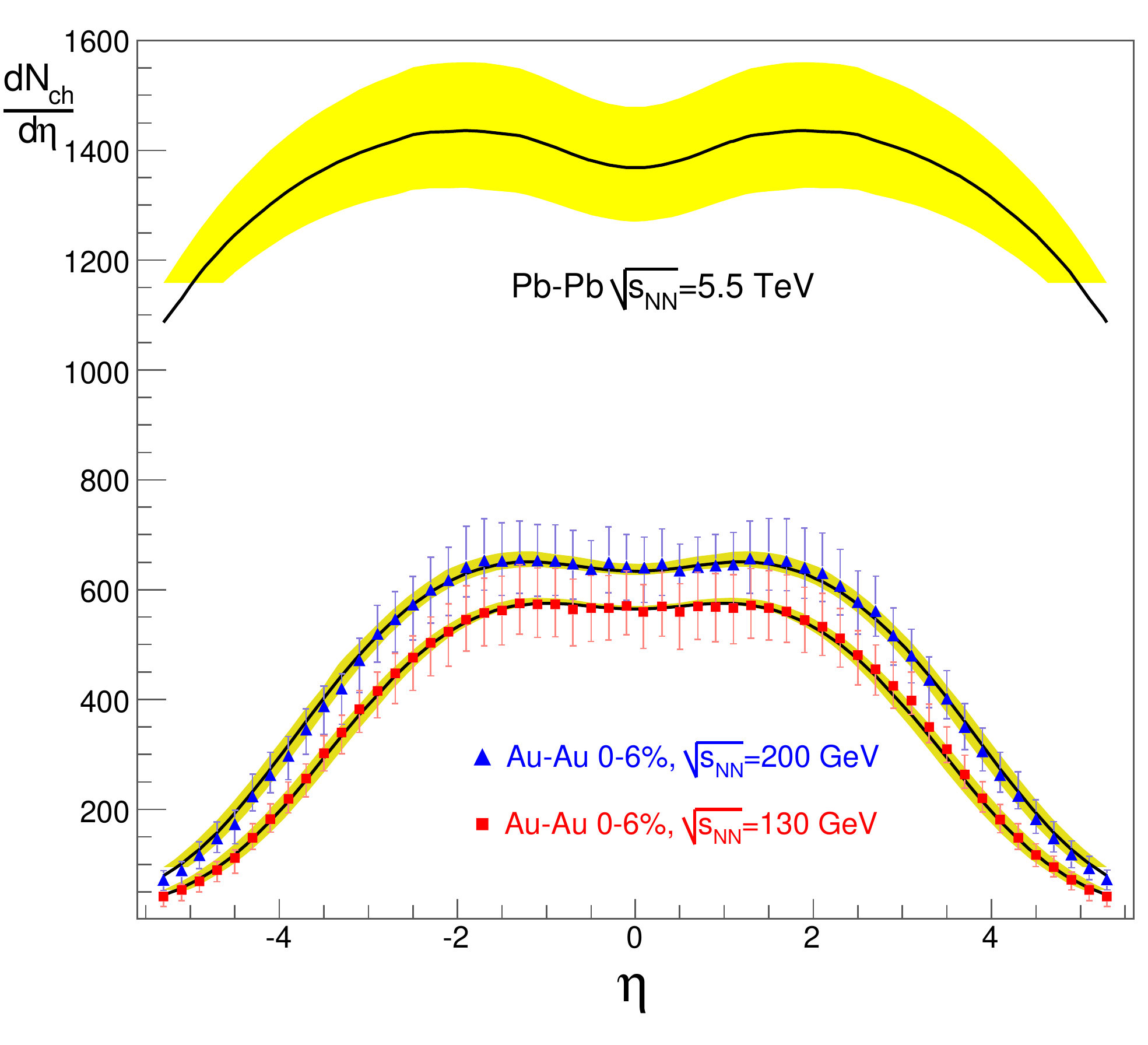,height=6.8cm}
\end{center}
\caption{Multiplicity densities as a functiion of pseudorapidity in RHIC Au+Au collisions at $\sqrt{s}=200$ and 130 GeV (data by the PHOBOS Coll.). Uppermost curves correspond to predictions for Pb+Pb collisions at the LHC ($\sqrt{s}=5.5$ TeV).}
\label{dndy}
\end{figure}

\section{Forward single inclusive spectra in p+p, d+Au and p+Pb collisions.}

The experimentally observed suppression of forward hadron yields in d+Au collisions compared to those measured in p+p collisions was predicted in CGC base calculations, albeit at a qualitative level\cite{Albacete:2003iq,Kharzeev:2003wz}. Thanks to the new theoretical tools available, now it is also possible to get a good quantitative description of such suppression \cite{Albacete:2010bs}. At forward rapidities, where the projectile is probed at large-$x$ and the target nucleus at small-$x$, single inclusive hadron production can be calculated as \cite{Dumitru:2005gt}:
\begin{eqnarray}
\frac{dN_h}{dy_h\,d^2p_t}=\frac{K}{(2\pi)^2}\sum_{q}\int_{x_F}^1\,\frac{dz}{z^2}\, \left[x_1f_{q\,/\,p}
(x_1,p_t^2)\,\tilde{N}_F\left(x_2,\frac{p_t}{z}\right)\,D_{h\,/\,q}(z,p_t^2)\right.\nonumber\\ 
+\left. x_1f_{g\,/\,p}(x_1,p_t^2)\,\tilde{N}_A\left(x_2,\frac{p_t}{z}\right)\,D_{h\,/\,g}(z,p_t^2)\right]
\label{hyb}\,,
\end{eqnarray}
where $p_t$ and $y_h$ are the transverse momentum and rapidity of the produced hadron, and
$f_{i/p}$ and $D_{h/i}$ refer to the parton distribution function of the incoming proton and to the fragmentation function respectively. The gluon distributions representing the target are given by the Fourier transforms of the dipole amplitude: $\tilde{N}_{F(A)}(x,k)=\int d^2{\bf r}\,e^{-i{\bf k}\cdot{\bf r}}\left[1-\mathcal{N}_{F(A)}(r,x)\right]$, where F(A) stands for fundamental (adjoint) representation. 

\begin{figure}
\vskip 2.5cm
\epsfig{figure=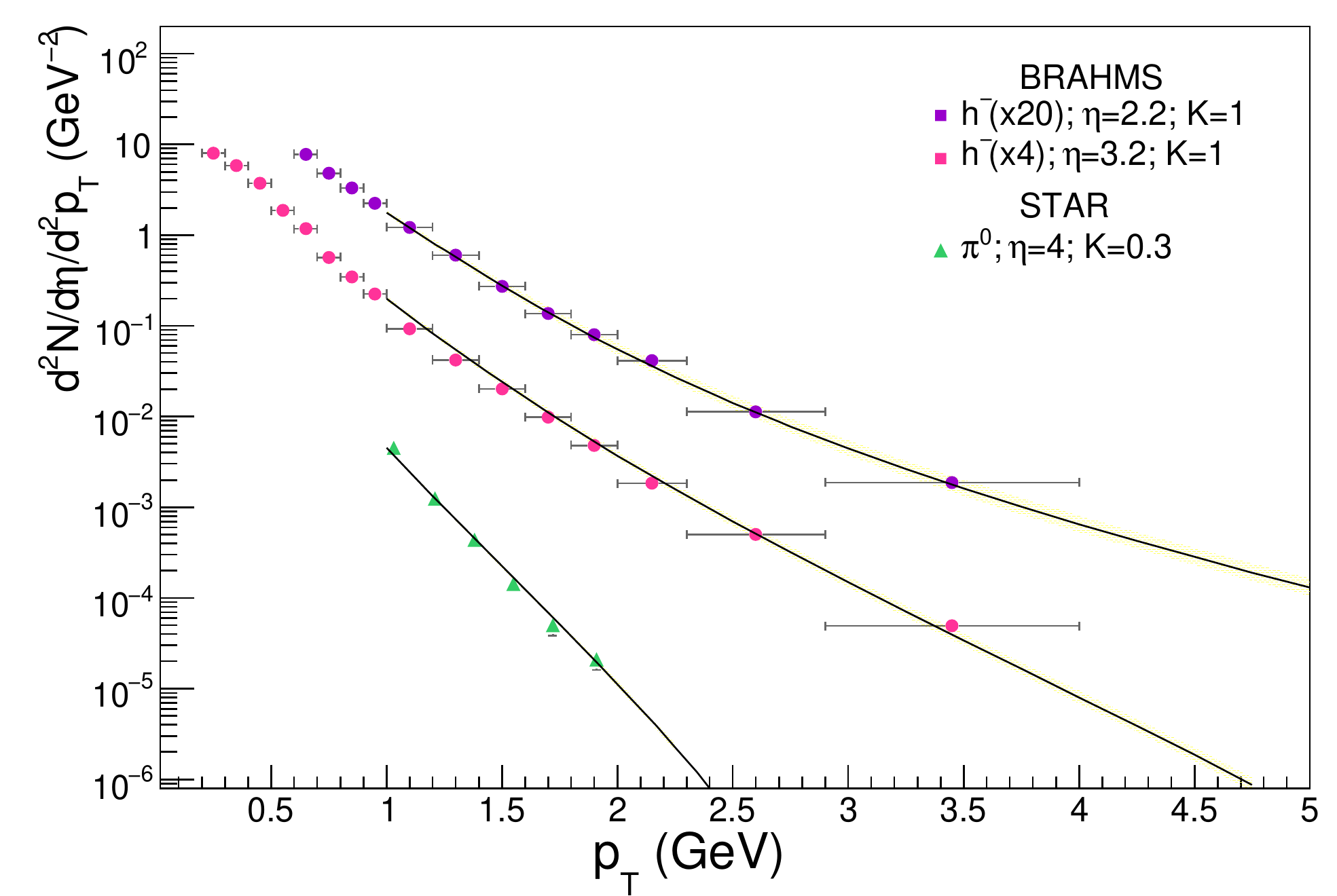,height=5.5cm}
\epsfig{figure=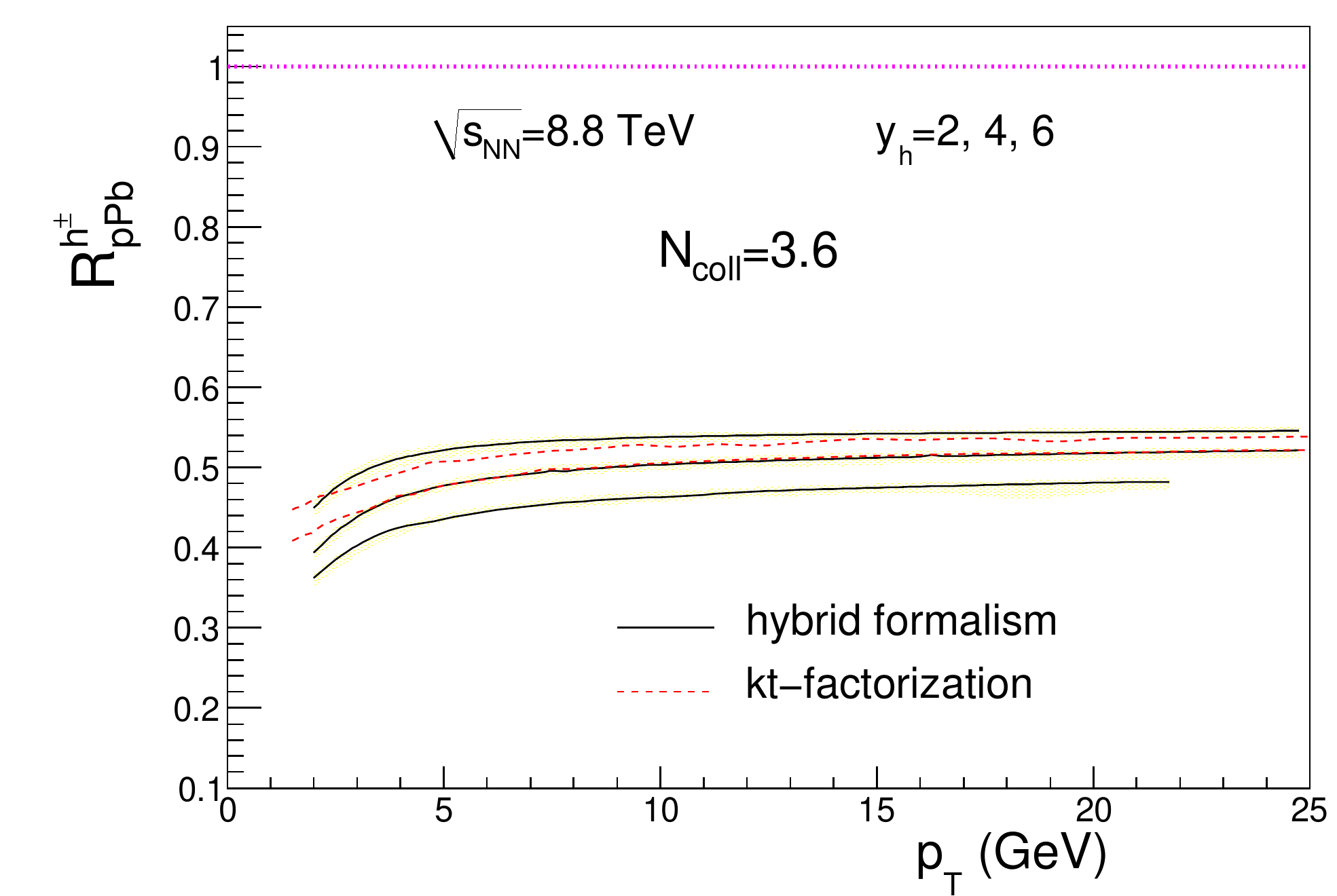,height=5.5cm}
\caption{Left: Charged hadron ($y\!=\!2.2$ and 3.2) and neutral pion ($y\!=4\!$) spectra in d+Au collisions. Right: Predicted nuclear modification factor for charged hadron spectra in p+Pb collisions at the LHC.}
\label{spec}
\end{figure}

With $Q_{s0}^2\!=0.4\div0.5$ GeV$^2$ (for quarks) and 0.005$<x_0<$0.025, we  \cite{Albacete:2010bs} obtain a very good description of the forward negative charged hadrons at $y\!=\!2.2$ and 3.2 (data by the BRAHMS Collaboration \cite{Arsene:2004ux}) and neutral pion production at $y\!=\!4$ (data by the STAR collaboration \cite{Adams:2006uz}) in minimum bias d+Au collisions, as shown in Fig. 2 (left). An equally good description is obtained for proton-proton data at the same collision energy with $Q_{s0}^2\!=0.2$ GeV$^2$. Remarkably, the description of negative charged hadrons does not require any $K$-factors, i.e. $K=1$. However, for neutral pion production we get $K=0.3\, (0.4)$ in d+Au (p+p) collisions. In agreement with other CGC calculations, we predict a sizable ($\sim0.5$) suppression for the nuclear modification factor $R_{pPb}=(dN_{h^\pm}^{pPb}/dyd^2p_t)/(dN_{h^\pm}^{pp}/dyd^2p_t)/N_{coll}$ where, following the experimental analyses at RHIC, we have taken $N_{coll}\!=\!3.6$. 

\section{Di-hadron azimuthal correlations at forward rapidities in d+Au collisions.}
Recent measurements of azimuthal forward di-hadron correlations in d+Au collisions by the STAR Collaboration \cite{Braidot:2010zh} exhibit the feature of {\it monojet} production, i.e., the suppression of the away-side peak characteristic of approximate back-to-back correlations. Following \cite{Marquet:2007vb}, we calculated the forward double inclusive pion production in d+Au collisions in the CGC framework \cite{Albacete:2010pg}. 
More specifically, we are interested in the coincidence probability, which is the experimental measured quantity. It is given by $CP(\Delta\phi)=N_{pair}(\Delta\phi)/N_{trig}$ with
\begin{equation}
N_{pair}(\Delta\phi)=\hspace{-0.4cm}\int\limits_{y_i,|p_{i\perp}|}\hspace{-0.3cm}
\frac{dN^{dAu\to h_1 h_2 X}}{d^3p_1 d^3p_2}\ ,\
N_{trig}=\hspace{-0.3cm}\int\limits_{y,\ p_\perp}\hspace{-0.2cm}\frac{dN^{dAu\to hX}}{d^3p}\ ,
\label{cp}
\end{equation}
and it has the meaning of the probability of, given a trigger hadron $h_1$ in a certain momentum range, produce an associated hadron $h_2$ in another momentum range and with a difference between the azimuthal angles of the two particles equal to $\Delta\phi$.
Following the experimental analysis by the  STAR collaboration \cite{Braidot:2010zh}, we set $|p_{1\perp}|>2$ GeV, $1\ \mbox{GeV}<|p_{2\perp}|<|p_{1\perp}|$ and $2.4< y_{1,2}< 4$ for the transverse momenta and rapidity of the produced pions. Our results, shifted by an arbitrary offset, are shown in Fig 3, together with the corresponding preliminary data by the STAR collaboration.
The disappearance of the away-side peak around $\Delta\phi\sim\pi$ in d+Au collisions exhibited by data is quantitatively well described by our CGC calculation. In our approach, the angular decorrelation is due to the the momentum broadening induced by the propagation of the projectile (or its Fock states) through the nucleus. The momentum broadening, in turn, is related to the saturation scale of the target nucleus, which is large in the forward region. Fig 3 (right) shows the centrality dependence of our calculation. We predict that the away peak should reappear for more peripheral collisions or, at a fixed rapidity, for larger momenta of the detected particles.

\begin{figure}
\begin{center}
\epsfig{figure=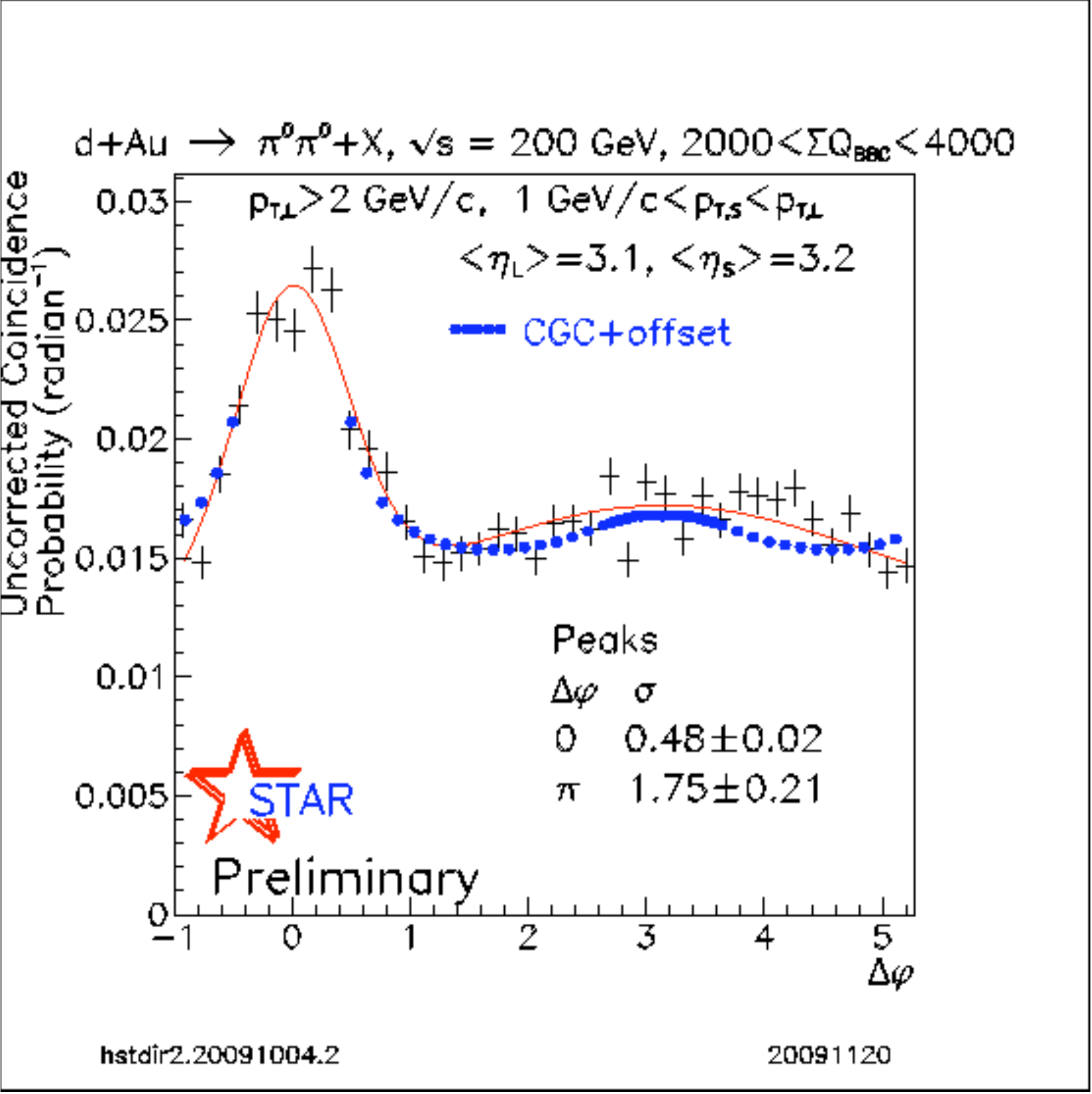,height=5.5cm}
\epsfig{figure=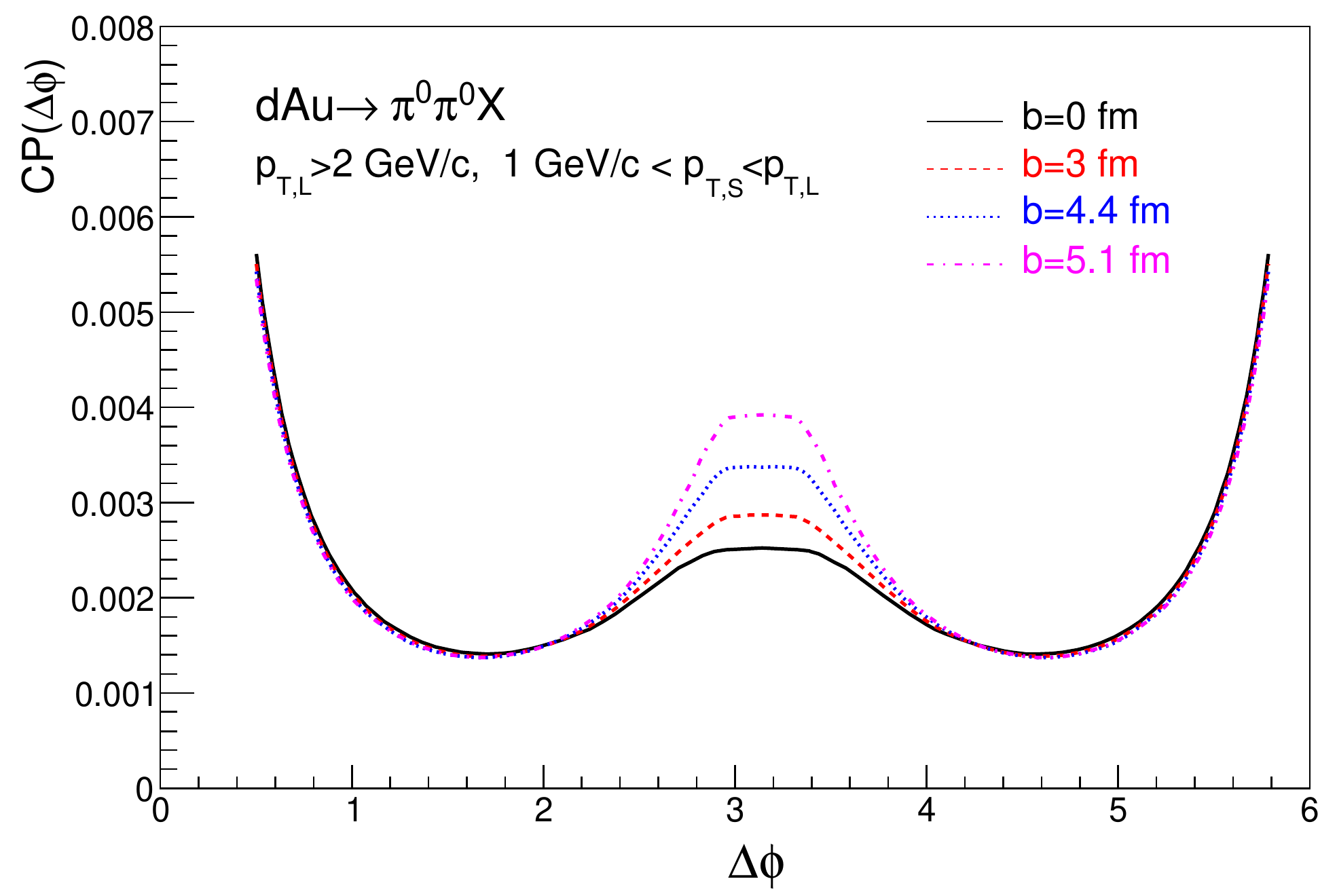,height=5.25cm}
\caption{Left: $CP(\Delta\phi)$ for forward pions in d+Au collisions (preliminary data by the STAR Coll.) and CGC theoretical results (blue dots). Right: $CP(\Delta\phi)$ for various collision centralities.}
\end{center}
\end{figure}

\section*{Acknowledgments}
This work is supported by a Marie Curie Intra-European Fellowship (FP7- PEOPLE-IEF-2008), contract number 236376.

\section*{References}

\begin{thebibliography}{18}

\bibitem{Gelis:2010nm}
F.~Gelis, E.~Iancu, J.~Jalilian-Marian, and R.~Venugopalan.
\newblock  arXiv:1002.0333 [hep-ph].

\bibitem{Balitsky:2006wa}
I.~I. Balitsky.
\newblock {\em Phys. Rev. D}, 75:014001, 2007.

\bibitem{Kovchegov:2006vj}
Y.~Kovchegov and H.~Weigert.
\newblock {\em Nucl. Phys. {\bf A}}, 784:188--226, 2007.

\bibitem{Gardi:2006rp}
  E.~Gardi, J.~Kuokkanen, K.~Rummukainen and H.~Weigert,
\newblock {Nucl.\ Phys.\  A {\bf 784} (2007) 282}

\bibitem{Albacete:2007yr}
Javier~L. Albacete and Yuri~V. Kovchegov.
\newblock {\em Phys. Rev.}, D75:125021, 2007.

\bibitem{Balitsky:1995ub}
  I.~Balitsky,
  Nucl.\ Phys.\  B {\bf 463}, 99 (1996).

\bibitem{Kovchegov:1999yj}
  Y.~V.~Kovchegov,
  Phys.\ Rev.\  D {\bf 60}, 034008 (1999).

\bibitem{Kharzeev:2001yq}
Dmitri Kharzeev, Eugene Levin, and Marzia Nardi.
\newblock {\em Phys. Rev.}, C71:054903, 2005.


\bibitem{Albacete:2007sm}
Javier~L. Albacete.
\newblock {\em Phys. Rev. Lett.}, 99:262301, 2007.

\bibitem{Albacete:2003iq}
Javier~L. Albacete, Nestor Armesto, Alex Kovner, Carlos~A. Salgado, and
  Urs~Achim Wiedemann.
\newblock {\em Phys. Rev. Lett.}, 92:082001, 2004.

\bibitem{Kharzeev:2003wz}
Dmitri Kharzeev, Yuri~V. Kovchegov, and Kirill Tuchin.
\newblock {\em Phys. Rev.}, D68:094013, 2003.


\bibitem{Albacete:2010bs}
Javier~L. Albacete and Cyrille Marquet.
\newblock {\em Phys. Lett.}, B687:174--179, 2010.

\bibitem{Dumitru:2005gt}
Adrian Dumitru, Arata Hayashigaki, and Jamal Jalilian-Marian.
\newblock {\em Nucl. Phys.}, A765:464--482, 2006.


\bibitem{Albacete:2010pg}
Javier~L. Albacete and Cyrille Marquet.
\newblock  arXiv:1005.4065 [hep-ph].


\bibitem{Braidot:2010zh}
Ermes Braidot for the~STAR collaboration.
\newblock{arXiv:1005.2378 [hep-ph]. In these proceedings.}


\bibitem{Marquet:2007vb}
Cyrille Marquet.
\newblock {\em Nucl. Phys.}, A796:41--60, 2007.

\bibitem{Arsene:2004ux}
  I.~Arsene {\it et al.}  [BRAHMS Collaboration],
  \newblock{Phys.\ Rev.\ Lett.\  {\bf 93} (2004) 242303}

\bibitem{Adams:2006uz}
  J.~Adams {\it et al.}  [STAR Collaboration],
  Phys.\ Rev.\ Lett.\  {\bf 97}, 152302 (2006)

\end{thebibliography}


\end{document}